\documentclass[10pt]{article}
\usepackage{amsmath,amssymb}

\begin{document}
\title{Discrete Newtonian Cosmology}
\author{Gary W Gibbons \\
Trinity College and DAMTP, Cambridge University,\\
\\
George F R Ellis,\\
ACGC and Department of Mathematics, University of Cape Town, \\
Trinity College and DAMTP, Cambridge University.}
\date{\today}
\maketitle

\begin{abstract}
In this paper we lay down the foundations for a purely Newtonian
theory of cosmology, valid at scales small compared with the Hubble radius,
using only Newtonian point particles acted on by gravity and a possible
cosmological term. We describe the cosmological
background which is given by an exact solution of the equations
of motion in which the particles expand homothetically
with their comoving positions constituting  a central configuration.
We point out, using previous
work, that an important class of central configurations
are homogeneous and isotropic, thus justifying
the usual assumptions of elementary treatments.
The scale factor is shown to satisfy the standard  Raychaudhuri
and Friedmann equations  without making  any fluid dynamic
or continuum approximations. Since we make no commitment as to the identity
of the point particles, our results are valid for cold dark matter, galaxies, or
clusters of galaxies.  In future publications we plan to
discuss perturbations of our cosmological background
from the point  particle viewpoint laid down in this paper
and show consistency with much standard theory
usually  obtained by  more complicated and conceptually
less clear continuum methods. Apart from its potential
use  in large scale structure studies, we believe
that our approach has great pedagogic advantages over
existing elementary treatments of the expanding universe, since it requires no use
of general relativity or continuum mechanics but concentrates
on the basic physics: Newton's laws for gravitationally interacting particles.

\end{abstract}

\section{Discrete Newtonian cosmology}
\label{sec:intro} It is customary in present day astrophysical cosmology
(e.g. \cite{Pee80,Pee93}) to assume that Newtonian theory describes
adequately what happens after the time of decoupling of matter and
radiation. This is particularly true of large scale structure formation
theories, which are all essentially based in Newtonian gravity
\cite{BagPad97,Ber98,TreHut08}. It is then an issue as to what is the best way to
express a Newtonian cosmological theory, given that Newton himself failed in
this endeavour \cite{Hoskin1,Hoskin2}.\newline

One reason for Newton's failure is that he adopted
a static model in which the fixed stars extended to infinity.
This leads, amongst other things, to problems with divergences
analogous to Olber's
paradox. One way to evade these problems is to modify Newton's law
at large distances such as suggested by Neumann \cite{Neumann} and by
Seeliger \cite{Seeliger1,Seeliger2,Seeliger3,Seeliger4}.
However one can construct a viable Newtonian cosmological theory from
scratch for a finite number  of
discrete gravitating particles interacting through
Newtonian gravitational attraction only, but possibly with the Newtonian
version of a cosmological constant. The homothetically  expanding
background solution is an exact solution of the Newtonian equations, and can be
linearised to get perturbation theory for such models, relevant to structure
formation calculations. The model has many advantages over the standard
approach: the issue is whether  it can be made realistic.

\subsection{The issues}
Hence: consider a Newtonian cosmology for a universe made up of discrete
gravitating  particles. We do not assume either general relativity or
fluid dynamics. We just use Newton's Laws of motion applied to a set of
gravitating particles, with the particle interactions given by Newton's law
of gravitational attraction. The features of this model are,

\begin{itemize}
\item \textbf{A set of gravitating particles imbedded in a vacuum - no fluid.%
} The usual cosmological ``fluid'' is very problematic because standard
fluid properties are derived for particles that only undergo short range
interactions (collisions). This is not in the least like collision-less
particles that only interact by long range (gravitational) forces; this
includes a `fluid' of galaxies, stars, or CDM particles. The dangers of not
properly representing long range forces include divergences and neglecting
non-local interactions.

A further point is numbers of particles. The key question is, what
is the size of averaging volume $V_{avg}$ that is supposed to give a
good fluid approximation? \cite{Ba200}. A typical gas has around
$10^{23}$ particles per cubic centimeter. If we regard galaxies as
`particles' of cosmic fluid, we have at most $10^{11}$ particles to
consider in the visible universe, and if it is clusters we consider
 we have many
less. Sticking
with galaxies to be conservative, if we assume the averaging scale
$L_{avg}$ is (1/100 th). of the Hubble radius, then $L_{avg} \simeq
4.6 \times 10^5$ Mpc, and the observable universe contains
$100^3=10^6$ such averaging volumes each containing $10^5$ galaxies.
If we take it to be (1/1000 th). of the Hubble radius $(L_{avg}
\simeq 4.6 \times 10^4$ Mpc), the observable universe contains
$1000^3=10^9$ averaging volumes each containing $10^2$ galaxies. In
both cases this  is much too small for a good fluid approximation. Now
of course there is dark matter present in addition, but that too is
clustered in mini-halos. It is not at all obvious there is a good
fluid description applicable at the cosmological scale.

What is desirable is to represent the discrete particles and their
interactions (if there is a good fluid approximation, this will
underly it; if not, this will be vastly preferable). Thus one
represents individual particles and uses summation rather than
integration.

\item \textbf{No infinities}: the basic reason Newton failed to get a
cosmological solution was the divergences associated with an assumed
infinite number of particles. One does not need to make this assumption: one
can do the calculation for a finite set of particles and avoid these
divergences, and all the associated problems \cite{Nor99,Malament,Layzer1,
Layzer2,McCrea1,McCrea2,Heckman1,Heckman2,Heckman3,Tipler}.

\item \textbf{No Fourier analysis}: for calculating the dynamics, one can
work with the actual distribution rather than its Fourier modes, and then
Fourier analyse the results afterwards if it is  helpful. Fourier analysis of the
dynamics is great if the system is linear; if it is not, the linearity
assumed in Fourier analysis won't work; and the basic gravitational
interactions for structure formation are non-linear (although we can of
course perturb to get linear solutions about a background model).

\item \textbf{No periodic boundary conditions} One can avoid the assumption
of periodic boundary conditions which is usually made to deal with the problem of
boundary conditions. This restricts the nature of solutions allowed, and may
introduce artefacts, unless the spacetime actually is periodic in the
assumed way. For example angular momentum will not always be conserved
because a system with periodic boundary conditions is not rotationally
symmetric (see \cite{ShiBurJoh06} for such effects in molecular dynamics
simulations).
\end{itemize}

The `particles' in question may be envisaged as stars, galaxies,
clusters, superclusters, etc., or even molecules. Indeed there is a
certain arbitrariness in the choice of what a particle is since any
sufficiently spherically symmetric isolated subsystem will move and
gravitate like a point particle located at its
``centre''.\\

This paper is the same in spirit as the general relativity paper by
Lindquist and Wheeler \cite{LinWhe57} and subsequent papers by Redmount,
Ferreira, Clifton, and others \cite{Redmount,CliFer09,Cli11,CliRosTav12,Yoo1,Yoo2}.
However those are all based on regular lattices, which will
presumably not be solutions of the central configuration equation. They are
also approximate rather than exact solutions.

\subsection{The outcome}
The outcome of this approach is that we if we have a suitable initial distribution of discrete particles, that is
one satisfying the \emph{central configuration equation} (see (\ref{Newt3})
below), we obtain an exact Newtonian version of the standard FLRW models ---
a solution that expands homothetically (equations (\ref{homothetic}) and (\ref%
{vel}) are satisfied) according to the standard Friedmann equation and
Raychaudhuri equations for pressure-free matter (equations (\ref{Fried2})
and(\ref{ray}) below), no matter how many particles are involved. For large
numbers of particles, the solutions are close to spatially homogeneous \cite{BatGibSut03}.\newline

The central configuration equation is in effect an initial value equation
for these models, and is \emph{required} if we are to have such a FLRW-like
model. We can then perturb about such a model in order to get Newtonian
equations of structure formation for such models; or else we can numerically
integrate to see what happens if the initial data is changed from the
background state in a linear or non-linear way.\newline

In a subsequent paper, this approach will be developed further, deriving the linearised Newtonian
structure formation equations in a rigorous way, deriving effects such as
the Zeldovich pancake models \cite{ShaZel89}, and hopefully providing a
basis of this kind for non-linear equations and numerical simulations. It
may be only of theoretical/didactic interest: but it might provide
a sound basis for looking again at N-body simulations \cite{BagPad97,Ber98,TreHut08}
in a cosmological context.

\subsection{This paper}
In the following section (Section \ref{sec:basic}), we set up the basic theory for a collection of point particles interacting only via the Newtonian inverse square gravitational law. We derive the generic forms of the various associated conservation laws, potential energies, and Lagrangian, as well as the general form of the virial relation. \\

Section \ref{sec:cosm} applies this theory to cosmology, showing how exact homothetically expanding solutions, with the same behaviour of their
dynamical  equations as their General Relativity counterparts, are possible provided the central configuration equation is satisfied. This result is summarised in a main theorem presented in section \ref{sec:main_result}. The result is unaffected by the existence of a cosmological constant (Section \ref{sec:cc}). Exact and expanding solutions exist as precise analogs of the General Relativity pressure-free cosmological solutions (Section \ref{sec:solutions}). The central configuration equation is key to these exact solutions of the Newtonian equations; Section \ref{sec:cce} studies properties of this equation, in particular looking at the related effective forces and potentials. \\

This paper sets the stage for various generalisations (Section \ref{sec:further}). A further paper that will consider properties of perturbed versions of these solutions, representing structure formation in an expanding universe.

\section{Basic Theory}
\label{sec:basic}

In this section we shall review some standard material
\cite{Pollard,Boccaletti, Arnold,Saari0} on the dynamics
of $N$ point particles moving under the influence of gravity which we shall need in the later part of the paper.

\subsection{The basic equations}
The basic equation for a set of gravitating masses only interacting amongst
themselves is Newton's Law of attraction. In general for interacting point
particles, on using inertial coordinates, Newton's force law for discrete
particles at position $\mathbf{x}_a$ and with mass $m_a > 0$ is
\begin{equation}  \label{general}
m_a \frac{d^2\mathbf{x}_a}{dt^2} = \mathbf{F}_a + \sum_{b \neq a}\mathbf{F}%
_{ab} \,,
\end{equation}
where $\mathbf{F}_a$ are external forces due to particles outside the set
considered, and $\mathbf{F}_{ab}$ inter-particle forces between particle $a$
and $b$. The same equation holds for each particle in the system, i.e. as $a$
ranges over the values $1,2,...,N$ if there are $N$ particles. The
gravitational force between any two particles is
\begin{equation}  \label{grav}
\mathbf{F}_{ab}^{(grav)} = - \frac{G m_a m_b } {|\mathbf{x}_a -\mathbf{x}%
_b|^3}(\mathbf{x}_a -\mathbf{x}_b) \,,
\end{equation}
where $G$ is Newton's gravitational constant. Thus for the gravitational
case,
\begin{equation}  \label{Newtplus}
m_a \frac{d^2\mathbf{x}_a}{dt^2} = -\sum_{b\neq a} G m_a m_b\frac{(\mathbf{x}%
_a-\mathbf{x}_b)}{|\mathbf{x}_a-\mathbf{x}_b|^3} + \mathbf{F}_a \,.
\end{equation}
The external forces vanish if

\begin{itemize}
\item We assume that the universe consists of a very large but finite number
of particles, and apply the force law to the entire set.

\item Or we apply (2) to a finite subset of all the particles in the
universe (which may or may not be finite) and assume that for reasons of
symmetry (most typically spherical symmetry, but not exclusively so) the
external force due to all the particles outside the subset cancel.
\end{itemize}

Thus if such forces cancel, or if we are considering all particles that
exist, then $\textbf{F}_a = 0$ and we get, for each $a$,
\begin{equation}  \label{Newt}
m_a \frac{d^2\mathbf{x}_a}{dt^2} = -\sum_{b\neq a} G m_a m_b\frac{(\mathbf{x}%
_a-\mathbf{x}_b)}{|\mathbf{x}_a-\mathbf{x}_b|^3}
\end{equation}
showing how
\begin{equation}  \label{Fgrav}
\mathbf{F}_a^{(grav)}
:= -\sum_{b\neq a} G m_a m_b\frac{(\mathbf{x}_a-%
\mathbf{x}_b)}{|\mathbf{x}_a-\mathbf{x}_b|^3}
\end{equation}
is the total gravitational force exerted on particle $a$ due to all the
other particles. Thus this is a coarse-grained or collective representation
of all the individual forces acting on the particle.

\subsection{Potential Energy}
The gravitational force $\mathbf{F}_a^{\rm grav}$ acting on the $a$-th
particle can be represented as the derivative of a gravitational potential
energy $V_a$ acting on that particle. \footnote{In contrast to many treatises
on celestial mechanics, but in accordance with the  universal  usage in physics
 the sign of the potential is chosen so that the
force it produces is in the direction in which the potential decreases.}
 The potential energy $V_a(\mathbf{x}_c)
$ for the gravitational force on the particle $\mathbf{x}_a$ is a function
of the position $\mathbf{x}_a$ defined by
\begin{equation}  \label{potdef1}
V_a(\mathbf{x}_a) := -\sum_{b \neq a} \frac{G m_a m_b}{|\mathbf{x}_a - \mathbf{x}%
_b|},
\end{equation}
which also depends on the positions $\mathbf{x}_c$ of all the other
particles in the system. This is the discrete version of the continuous
definition of this potential (see Saslaw \cite{Sas87}: equations (9.4) and
(9.5)).\footnote{%
Saslaw \cite{Sas87} shifts $m_a$ to the force relation (9.3). He also uses a different
sign for the potential.}\newline

To show the relation of this potential to the gravitational force, define
\begin{equation}
\mathbf{x}_{ba}:=\mathbf{x}_{b}-\mathbf{x}_{a},\,\,x_{ba}:=|\mathbf{x}%
_{ba}|=\left( (\mathbf{x}_{b}-\mathbf{x}_{a}).(\mathbf{x}_{b}-\mathbf{x}%
_{a})\right) ^{1/2}.  \label{def}
\end{equation}
From these definitions, for $\mathbf{x}_{a} \neq \mathbf{x}_{b}$,
\begin{eqnarray}
\frac{\partial }{\partial \mathbf{x}_{a}}(\frac{1}{x_{ba}})
&=&(\frac{1}{x_{ba}})^{3}(\mathbf{x}_{b}-\mathbf{x}_{a})  \label{dd}
\end{eqnarray}
where the partial derivative $(\partial/\partial \mathbf{x}_{a})$ is taken
keeping all the other positions $\mathbf{x}_{b}\,\,(b \neq a)$ constant. On using this result together with
 (\ref{grav}) and (\ref{Fgrav}),
the relation of the  potential (\ref{potdef1}) to the gravitational force
is
\begin{eqnarray}
\frac{\partial V_a}{\partial \mathbf{x}_{a}} &=&-\sum_{b \neq a}\mathbf{F}%
_{ab}^{(grav)}=-\mathbf{F}_{a}^{(grav)}  \label{pot1}
\end{eqnarray}
as required. As usual, the absolute value of the potential does not affect
this relation; we could add a constant $V_0$ to (\ref{potdef1}) without
affecting the result. However doing so would destroy an important property
of the potential: it is a homogeneous function of degree -1. This play an
important role in various relations (sections \ref{sec:virial} and \ref%
{sec:idneitity}).

\subsection{Symmetries of the equations, and of the solutions}

The fundamental equation  (\ref{Newt}) incorporates the important
physical  property that inertial, and both active and passive
gravitational masses are all equal. As a consequence
(\ref{Newt}) is invariant under the full ten-dimensional
Galilei group, the non-relativistic limit of the Poincar\'e  group.
\begin{itemize}
\item time translations ($t \rightarrow t + t_0$),
\item spatial translations ($\mathbf{x}_a \rightarrow \mathbf{x}_a + \mathbf{x%
}_0$ for all $a$),
\item rotations about the origin (no preferred direction is implied by the
vectors, and rotations preserve $|\mathbf{x}_a-\mathbf{x}_b|$),
\item Boosts from one frame of reference or inertial coordinate system
 to another.
\end{itemize}
The symmetries of the equations will generally not be symmetries of the
solutions. However they can be used to generate new solutions that are
essentially identical to the old ones.\\

In more detail:

\begin{itemize}
\item  {\bf 1: Translational  Invariance}
  As we are using a discrete model,
there will not be any continuous spatial invariance of the solutions.
Additionally, the homothetic  solutions we shall describe  have a preferred
barycentre, i.e. a preferred  centre of mass.

However:

\noindent (i) new solutions can be obtained from the old by
spatially translating them. These are physically identical to the old ones.
This can therefore be regarded as a change of coordinates (one is referring
the same physical system to a new coordinate system).

\noindent(ii) if there is a
large enough number of particles, the system will appear approximately
spatially homogeneous when coarse grained . Nevertheless if it is a finite
system it will have a boundary and so will not be spatially homogeneous on
a large scale.

Momentum conservation follows from translational  invariance of the equations.
This will be an exact result for the solutions even though they are not
spatially invariant.

\item{\bf 2. Rotational symmetry.}
Essentially the same remarks apply to rotational
symmetry. There cannot be continuous rotational symmetry because of the
discreteness of the system, but there can be discrete rotational
symmetries. With enough particles the solution will be approximately
rotationally symmetric when coarse grained. New physically identical
solutions can be generated by rotation of the old system.

Angular momentum conservation follows from rotational invariance of the
equations. This will be an exact result for the solutions even though they
are not rotationally invariant.

\item{\bf 3: Time symmetry invariance}
The solutions are only time invariant, i.e. invariant under shift of origin
of time, if static (with the scale factor $S(t) =$ constant) or
stationary ($S(t) = \exp Ht$).

Energy conservation results from time  invariance of the equations, and
will hold in all cases  unless external forces act on the system.

\item{\bf 4: Time reversal invariance}
There will be time reversal invariance of the solutions only if they are
static.  \footnote{It is striking that in  most if not all  early discussions, such as those of Newton and
his contemporaries \cite{Hoskin1,Hoskin2}, the concern was  that gravity would cause a finite system at rest to \emph{collapse}. But one should ask how the system could have  got to a starting point of being instantaneously 
at rest at a finite size at some time $t_0$, from which this collapse can occur.  
Unless one assumes existence of a cosmological constant giving such a static state, in which case expansion or collapse from that state are equally likely, this could only have happened by expansion from zero size at a 
previous time $t_i<t_0$ to that finite size. 
So that starting condition, assumed in their studies, 
implies the possibility of an expanding universe. If this had been realised at the beginning of the 20th century, cosmologists could have constructed a purely  Newtonian  explanation for the recession of the galaxies.}

If $\Lambda = 0$ there will be no static cosmological solutions, so they
will be non-static and there will necessarily be a start to the universe
($S = 0$ in the limit) where we can set $t = 0$ in the limit. In this case
we can choose time to be positive as the universe expands from this initial
singularity, so that $\dot{S} > 0$ for small enough $t >0$.  The direction
of time is then the direction in which $t$ increases near $t = 0$; it is
defined non-locally by the fact it is the direction in which the universe
initially was expanding (this is unaffected by whether the universe
recollapses or not). Thus a direction of time is established by this
cosmological context despite the time symmetry of the equations.

The local arrow of time will agree with this direction of time  if initial
conditions are special (see \cite{Ell13} for a discussion).

One can make the system appear to be dissipative by a choice of time parameter
non-linearly related to $t$, for example $\tau = \log\, t$ \cite{Barbour et al
2013}. However this is an artefact of an unphysical choice of the time
parameter (on the same basis, the simple harmonic oscillator will also appear
to be dissipative), and one would make the system appear to be dissipative in
the opposite direction of time if one instead chose $\tau = - \log \,t$. Standard
conservative Newtonian gravitational dynamics, and indeed standard physics in
general, results only if one restricts oneself to affine transformations of
the standard time function $t$.
\end{itemize}

There is in addition  a homothetic time symmetry:
\begin{equation}  \label{conformal1111} Equation
(\mathsf{\ref{Newt}) \,is\, invariant\, if\, } \qquad  (t \rightarrow At,\, \mathbf{x}%
_a \rightarrow A^{-\frac{2}{3} } \mathbf{x}_a) \forall \,A,\,a .
\end{equation}
This is an invariance of the basic equation and gives rise to a general form
of Kepler's third law in the form that if $\mathbf{x}_a(t)$
is a solution of the equations of motion then so
is $A^{-\frac{2}{3} } \mathbf{x}_a(At)$.
Of particular interest are solutions which are invariant under (\ref{conformal1111}):
\begin{equation}
\mathbf{x}_a(t) = A^{-\frac{2}{3} } \mathbf{x}_a(At) \,.
\end{equation}
For instance an important class of such solutions which we shall encounter
later take the form
\begin{equation}\label{invcon}
\mathbf{x}_a(t) =t^{\frac{2}{3}} \mathbf{r}_a \,,
\end{equation}
where $\mathbf{r}_a $ are independent of time.

\subsubsection{Mass and momentum and angular momentum  conservation}
The equality of active and passive
gravitational masses and the central nature
of the gravitational  force  guarantee not only
the conservation of momentum, angular momentum and energy but
also  the so-called Centre of Mass theorem, namely (i) the centre of mass, or barycentre, of an isolated system   moves with constant velocity and
(ii) one may always pass to a frame of reference by means of a Galilean transformation, i.e. a boost,  with respect which the centre of mass is at rest. This does not follow from the Galilei invariance alone \cite{Treder3}.\\

We assume particle mass is conserved:
\begin{equation}  \label{mass_cons}
dm_a/dt = 0.
\end{equation}
It then follows from (\ref{mass_cons}) and the symmetries of (\ref{Newt})
that total mass $M$, momentum $\mathbf{P}$, and angular momentum $\mathbf{L}$
about the origin are conserved:
\begin{eqnarray}
M & =& \sum_a m_a = M_0 \, (\mathrm{constant}) > 0,  \label{mass} \\
\mathbf{P} &=& \sum_a m_a \dot{\mathbf{x}}_a = \mathbf{P}_0 \,(\mathrm{%
constant}),  \label{mom} \\
\mathbf{L} &=& \sum_a m_a (\mathbf{x}_a \times \,\dot{\mathbf{x}}_a) =
\mathbf{L}_0 \,(\mathrm{constant}) .  \label{angmom}
\end{eqnarray}
If $G$ was a function of time: $G = G(t)$, both $\mathbf{P}$ and $\mathbf{L}$
would still be conserved.
The conservation of momentum implies that the centre of mass
moves with constant velocity and that a frame of reference, i.e.
a set of inertial   coordinates,
may always be chosen so that the total momentum vanishes and the
centre of mass is at rest at the origin.
In what follows, this choice will always be made, unless stated otherwise.
Since the total momentum  ${\bf P}$,  total angular momentum ${\bf L}$ and
total  energy ${\cal E}$ , depend on the frame of reference, unless
stated otherwise, in what follows these quantities will be
taken with respect to  the centre of mass frame.

\subsubsection{Energy conservation}

Additionally energy $\mathcal{E}$ is conserved. By standard arguments

\begin{equation}
\mathcal{E} = T + V = \mathcal{E}_0\, (\mathrm{constant})\,,  \label{energy}
\end{equation}
where the kinetic energy $T$ and potential energy $V$ are
\begin{eqnarray}  \label{energydef}
T(\dot{\mathbf{x}_c}) &:=& \frac{1}{2}\sum_a m_a (\dot{\mathbf{x}}_a)^2, \\
V(\mathbf{x}_c) &:=& \sum_a V_a = -\sum_{a}\sum_{b \neq a} \frac{G
m_a m_b}{| \mathbf{x}_a - \mathbf{x}_b|}  \label{potdef}
\,.\end{eqnarray}
The total gravitational potential energy,
$V(\mathbf{x}_c)$ is  homogeneous  $(V(a\mathbf{x}) = a^k V(%
\mathbf{x})$) of degree $k = -1$ . It is a negative function of the set $\{%
\mathbf{x}_c\}$ of the positions of all the particles in the system, and
gets more negative the closer they are together. $V(\mathbf{x}_c)$ is thus a
function of $3N$ coordinates $\mathbf{x}_c$ ($1\leq c \leq N$) of the $3N$
dimensional configuration space $Q$ of $N$ points in $R^3$, and is a continuous
and indeed analytic function of these coordinates away from the diagonal
where two or more positions coincide. \newline

The quantities $T$ and $V$ are numbers that are both coarse grained
representations of the state of the system, $T$ representing the total
energy of motion of the particles, and $V$ the sum of the potential energies
of all the particles. The gradient of $V$ does not represent any force;
indeed as it is just a number, it is not a function which can have a spatial
gradient.\footnote{%
Potential energy in an external gravitational field is a function of
position; in that case the external force $\mathbf{F}_a$ is non-zero.}

\subsection{Moment of inertia and virial theorem}

Half the moment of inertia about the centre is \footnote{The reader is warned that
many books on celestial  mechanics define $I$ without the factor
of $\frac{1}{2}$. This leads to various differences with the  formulae
in books which do not use our convention.}
\begin{equation}
I=\frac{1}{2}\sum_{a}m_{a}\mathbf{x}_{a}.\mathbf{x}_{a}=\frac{1}{2}%
\sum_{a}m_{a}x_{a}^{2}  \label{cofm}
\end{equation}%
which plays an important role in celestial dynamics. The quantity $I^{1/2}$ serves as a measure of the maximum spacing of
particles, while $V^{-1}$ serves as a measure of their minimum spacing \cite{Mar76}. 
Sundman's inequality is
\begin{equation}
(\mathbf{L}_{0})^{2}+(dI/dt)^{2}\leq 4IT.  \label{Sund}
\end{equation}%
It  plays  a role in Newtonian non-singularity theorems stating that no
complete collapse (i.e. one for which $I \rightarrow 0$ ) must occur
in finite time,  and cannot  occur if $\mathbf{L}_{0}^{2} \neq 0$
\cite{Pollard,Barrow}.

\subsubsection{The virial theorem}

This is a standard result which depends  crucially on the scaling property
of the potential energy, i.e on Newton's inverse square law.
\label{sec:virial} Take a dot product of $\mathbf{F}_{a}^{(grav)}$ given by (%
\ref{Fgrav}) with $\mathbf{x}_{a}$, and sum over $a$ to get
\begin{equation}
\sum_{a}\mathbf{x}_{a}.\mathbf{F}_{a}^{(grav)}=-\sum_{a}\sum_{b\neq
a}Gm_{a}m_{b}\frac{\mathbf{x}_{a}.(\mathbf{x}_{a}-\mathbf{x}_{b})}{|\mathbf{x%
}_{a}-\mathbf{x}_{b}|^{3}}=\sum_{a}\sum_{b\neq a}\mathbf{x}_{a}.\partial _{%
\mathbf{x}_{a}}(\frac{Gm_{a}m_{b}}{|\mathbf{x}_{a}-\mathbf{x}_{b}|})
\,.\label{virrr}
\end{equation}%
Now Euler's theorem on homogeneous functions of degree $k$ (that is
functions $f(V)$ such that $f(ax)=a^{k}f(x)$) says
\begin{equation}
x\,\partial f/\partial x=kf  \label{euler} \,.
\end{equation}%
In this case $f=\frac{1}{|\mathbf{x}_{a}-\mathbf{x}_{b}|}$ is of degree $k=-1
$, so Euler's theorem says
\begin{equation}
\mathbf{x}_{a}.\partial _{\mathbf{x}_{a}}f=-f\Rightarrow \mathbf{x}%
_{a}.\partial _{\mathbf{x}_{a}}(\frac{1}{|\mathbf{x}_{a}-\mathbf{x}_{b}|})=-%
\frac{1}{|\mathbf{x}_{a}-\mathbf{x}_{b}|}.  \label{euler1} \,.
\end{equation}%
By (\ref{potdef}) the last term of (\ref{virrr}) is
\begin{equation}
\sum_{a}\sum_{a\neq b}\mathbf{x}_{a}.\partial _{\mathbf{x}_{a}}(\frac{%
Gm_{a}m_{b}}{|\mathbf{x}_{a}-\mathbf{x}_{b}|})=-\sum_{a}\sum_{b\neq a}(\frac{%
Gm_{a}m_{b}}{|\mathbf{x}_{a}-\mathbf{x}_{b}|})=V \,. \label{Newt32111}
\end{equation}%
Also from the equation of motion (\ref{Newt}),
\begin{eqnarray}
\sum_{a}\mathbf{x}_{a}.\mathbf{F}_{a}^{(grav)} &=&\sum_{a}m_{a}\mathbf{x}%
_{a}.\frac{d^{2}\mathbf{x}_{a}}{dt^{2}}=\sum_{a}m_{a}(\frac{d}{dt}(\mathbf{x}%
_{a}.\frac{d\mathbf{x}_{a}}{dt})-\frac{d\mathbf{x}_{a}}{dt}.\frac{d\mathbf{x}%
_{a}}{dt})  \nonumber \\
&=&\frac{d}{dt}\sum_{a}m_{a}\frac{1}{2}\frac{d}{dt}(\mathbf{x}_{a}.\mathbf{x}%
_{a})-2T  \label{viriall111}
\end{eqnarray}%
and so from equation (\ref{virrr}) with (\ref{viriall111}) and using (\ref%
{Newt32111}) and (\ref{cofm}) we find
\begin{equation}
V=\frac{d^{2}I}{dt^{2}}-2T  \label{virialrelation}
\end{equation}%
which is the scalar virial equation (\cite{Sas87}: eqn.(9.16)).
In the celestial mechanics literature
equation (\ref{virialrelation}) is called the  \emph{Lagrange-Jacobi equation}.
\newline

Taking a time average $\langle\,\,\rangle$ of this equation, if the average of the second derivative of $I(t)$ is zero, we get a relation between kinetic and potential energies:
\begin{equation}  \label{virialthm}
\langle \frac{d^{2}I}{dt^{2}}\rangle = 0 \,\,\Rightarrow\,\, \langle
V\rangle=-2\langle T\rangle
\end{equation}
which is the virial theorem (\cite{Sas87}: pp.49-53). The condition $\langle
\frac{d^{2}I}{dt^{2}}\rangle = 0$ will be true for suitably localized or periodic
systems.

\subsection{Lagrangians and taking out the centre of mass}

The basic equation (\ref{Newtlamb}) in Section 3.2 below, which is (\ref{Newt}) with inclusion of a cosmological constant,
may be derived from the Lagrangian
\begin{equation}
L= \sum_a \frac{1}{2} m_a \dot {\bf x}_a ^2+ \frac{1}{2} \sum_{a \ne b} \sum _b
G {m_a m_b \over  |  {\bf x}_a-{\bf x}_b  | } + {1 \over 2 \tau^2}
\sum _a m_a  {\bf x} _a ^2  \,,
\end{equation}
with $\Lambda = 3 \frac{1}{\tau^2}$.
If we define

\begin{equation}
M= \sum_a m_a \,,\qquad  {\bf X} = {1 \over M} \sum _a m_a {\bf x}_a
\end{equation}
we find that
\begin{equation}
\ddot {\bf X} = {1 \over \tau^2 } {\bf X}
\end{equation}
so that
\begin{equation}
{\bf  X} = {\bf a} \cosh {t \over \tau} +{ \tau \bf u  } \sinh{ t \over \tau}
\end{equation}
where ${\bf a}$ and ${\bf u}$ are arbitrary constant vectors
with the dimensions of length and velocity respectively.
Now if $\{{\bf x}_a\}$ are solutions of the equations
of motion, then so are $\{{\bf x}_a+ {\bf X} \}$, and in fact
the Lagrangian $L$ is easily seen to change under this transformation by a total derivative:
\begin{equation}
L \rightarrow L+ \dot F
\end{equation}
with
\begin{equation}
\dot F = {3 \over 2}M \bigl(( {{\bf a} ^2 \over \tau ^2}  + {\bf u}^2  )
 \cosh(2 {t\over \tau})
+ 2 { {\bf a}.{\bf u} \over \tau} \sinh ( 2{t \over \tau})        \bigr)
\end{equation}
It follows that the equations of motion are invariant
not only under the three dimensional  group of  rotations $SO(3)$, but
under the six-dimensional commutative group of translations and boosts.
If we suppose that $G$ , which in principle could depend upon time,
is actually independent of time, then,  the equations of motion
are invariant under the ten-dimensional Newton-Hooke group.
which may be regarded as a deformation of the Galileo group
to which it tends as $\tau \rightarrow \infty$.  Note that the Newton-Hooke
group differs from the Galileo group in that  time translations
commute neither with boosts nor translations.\\

Because of this  symmetry of the equations of motion,
one would expect that one could obtain a formulation
which makes the translation  symmetry manifest (cf. \cite{Zanstra,Ding2,Ding3}). Such a formulation
is referred to as {\it relational}. To obtain it,
we subtract $ m_a \ddot {\bf X}$ from both sides of (\ref{Newtlamb})
 to obtain
\begin{equation}
{1 \over M} \sum _{b\ne a}  m_a m_b ( \ddot{\bf x} _a-\ddot {\bf x}_b )
=  -G \sum _{b \ne a}  m_a m_b { ({\bf x}_a-{\bf x}_b)
\over |  {\bf x}_a-{\bf x}_b  | ^3 } + {1 \over M \tau ^2} \sum_{b \ne a}
m_am_b ({\bf x}_a -{\bf x}_b ) \,.
\end{equation}

These equations may be obtained from the Newton-Hooke analogue
of Lynden-Bell's  reduced  so-called relational
Lagrangian \cite{Ding2}
\begin{equation}
L^\star = \sum _{\{a,b |a <b \}} {m_am_b \over M}  \Bigl \{
\frac{1}{2}   (\dot {\bf x} _a - \dot {\bf x}_b )^2 +
{ G M  \over |{\bf x}_a -{\bf x}_b  | } +
 { ( {\bf x}_a -{\bf x}_b ) ^2 \over 2 \tau ^2 }  \Bigr \}.
\end{equation}


\section{Cosmological solutions}\label{sec:cosm}

In this section we shall specialise the discussion
of the previous section to the case of solutions of the
Newtonian equations of motion which evolve by homotheties \footnote{Homotheties,
that is a constant rescaling of the Cartesian  coordinates
of Euclidean space, are also known as dilations, dilatations, similarities
 or homographies.}
of Euclidean space. Solutions of this type go back to
the work of Lagrange and Laplace  and are well known to those
studying celestial mechanics. Their  application to Newtonian
cosmology is much less well known. \\

Although the basic idea
that  homothetic expansion embodies the Cosmological Principle
is well explained  in chapter IX of  Bondi's 1952 textbook \cite{Bondi1},
Bondi  then goes on to review  Milne and McCrea's 1934
work \cite{Milne,MilneMcCrea} \footnote{Although often cited as the
initiators  of Newtonian Cosmology, Milne and McCrae
were preceded in 1932 by Mason \cite{Mason}, but his paper
appears to have received very little attention.} and treats  a  greatly simplified spherically symmetric
and homogeneous fluid dynamical model, apparently being unaware
that he could have equally well treat
 a fully rigourous point particle model,
from which both these two assumptions can be derived
rather easily\cite{BatGibSut03,GibPat03}. The key idea is that
Newton's  equations of motion for point particles only allow
a homothetic expansion if the co-moving  positions of  the
expanding system of particles are constrained  to  form what is called a
{\it central configuration}, as explained below.  As far as we are aware, the only papers
before  \cite{BatGibSut03} which made an explicit
link between central configurations
and Newtonian  Cosmology date back to 1971 and are by Saari
\cite{SaariApJ,Sar71,Mar76,Sar80,Saari0}  \footnote{However
the basic idea of substituting the homothetic ansatz into (\ref{Newt})
or (\ref{Newtlamb}) did occur to Landsberg \cite{Land1,Land1a}
slightly later although ; without tackling the central configuration equation.
Later \cite{Land2} he took up that challenge
but perhaps because he was only able to deal with a universe made of
eight equal mass galaxies situated at the vertices of a regular cube, his work
was not taken up at the time. The germ of the idea appeared in
two undergraduate textbooks \cite{Land3,D'Inverno}. Central configurations
with just four galaxies have been used to study the interactions between
nearby  galaxies \cite{Ding}.}
 whose primary
interest was mathematical celestial mechanics  rather than cosmology.
Moreover Saari's writings on central configurations is basically restricted to dealing only with a handful of particles. \\

Indeed most if not all of the celestial mechanics community
did not develop the essential physical insight based on  J. J. Thomson's
{\it current bun model of the atom}, and E.P. Wigner's
{\it Theory of Jellium}, an insight which can only be confirmed
by detailed  numerical analysis of the equations governing
the central configurations of ten   thousand particles or more \footnote{For Saari's most recent viewpoint on the importance of central configurations
for celestial mechanics see \cite{Sar21}.}.
This was the essential point of the work in \cite{BatGibSut03}.
This made use of the fact that  central configurations
extremise a certain function of position we call $\tilde V$.
In  \cite{BatGibSut03} it
 was established, in the case of  $N$  particles of equal mass $m$,
that for sufficiently many particles,  central configurations
which maximise $\tilde V$ form a spherical and homogeneous ball. This is  precisely the starting point of the
analysis of Milne and McCrea  \cite{Milne,MilneMcCrea}.
The homogeneity of central configurations made up of large numbers
of particles with identical or almost identical masses is a direct
consequence of Newton's inverse square law. Central configurations
exist for forces which vary inversely as any power but only for
the inverse square law  are they homogeneous \cite{BatGibSut03,Battye:2002ex}.


\subsection{Robertson-Walker-like solutions}
\label{sec:frw}

Now assume self-similarity of the solution: put in a homothetic factor $S(t)$
and separate variables so as to correspond to Robertson-Walker kinematics:
\begin{equation}
\mathbf{x}_a = S(t)\mathbf{r}_a,\,\,d\mathbf{r}_a/dt= 0, \label{homothetic}
\end{equation}
where $\mathbf{r}_a$ are comoving coordinates for particle $a$. This implies
the Slipher-Lema\^{\i}tre-Hubble velocity-distance law:
\begin{equation}  \label{vel}
\mathbf{v}_a := \frac{d\mathbf{x}_a}{dt} = \dot{S}(t)\mathbf{r}_a = H(t)%
\mathbf{x}_a
\end{equation}
where $H(t):=\dot{S}(t)/S(t)$. One should note here that the particle at the origin is moving inertially; only if $H(t) = H_0 t$ are the other particles also moving inertially. Thus except in this case (the Milne universe, which cannot occur for particles with non-zero masses), the origin is kinematically preferred. However, as pointed out by Heckmann and Sch\"{u}cking \cite{Heckman2,Heckman3}, we can use Einstein's insight that gravity and inertia are dynamically indistinguishable when accelerated motion occurs, so what counts physically is whether or not the particles are freely falling, that is, moving only under gravity plus inertia; and in that sense all the particles are dynamically equivalent.\footnote{See Section 2.3 in \cite{Ell71}.} No local dynamical effect will distinguish one particle from another.\\

The gravitational law (\ref{Newt}) becomes
\begin{equation}  \label{Newt1}
m_a \mathbf{r}_a \frac{d^2S(t) }{dt^2} =-\sum_{b\neq a} G m_a m_b\frac{(%
\mathbf{r}_a-\mathbf{r}_b)}{S^2(t)|\mathbf{r}_a-\mathbf{r}_b|^3}.
\end{equation}
Define
\begin{equation}  \label{defc}
C(t) := S^2(t)\frac{d^2S(t)}{dt^2}
\end{equation}
then equation (\ref{Newt1}) becomes
\begin{equation}  \label{Newt2}
C(t) m_a \mathbf{r}_a =-\sum_{b\neq a} G m_a m_b\frac{(\mathbf{r}_a-\mathbf{r%
}_b)}{|\mathbf{r}_a-\mathbf{r}_b|^3}.
\end{equation}
Then, remembering (\ref{mass_cons}) and (\ref{homothetic}), consistency
requires that $C(t)$ is a constant:
\begin{equation}  \label{cosnsistency}
\frac{\partial }{\partial t} \left(C(t) m_a \mathbf{r}_a\right) = 0
\Rightarrow C(t) = \mathrm{const} =: -G\tilde{M},
\end{equation}
which defines the constant $\tilde{M}$. Note that $\tilde M$ has the dimensions
of mass per unit volume.

\subsubsection{The central configuration  equation}

So firstly, from (\ref{Newt2}) together with (\ref{cosnsistency}) we must
have
\begin{equation}  \label{Newt3}
\tilde{M} m_a \mathbf{r}_a = \sum_{b\neq a} m_a m_b\frac{(\mathbf{r}_a-%
\mathbf{r}_b)}{|\mathbf{r}_a-\mathbf{r}_b|^3}
\end{equation}
for all values $a$. This set of $N$ non-linear time independent equations is
known as the \emph{central configuration equation} (\cite{BatGibSut03}; \cite{Arnold}:79-80), which
is a consistency condition for (\ref{homothetic}) to give a solution.
For systems with just a few particles, solutions form regular polyhedra.
This is clear for example in the case of three particles of identical mass:
if they are started off from rest in the shape of an equilateral triangle,
that shape  will be preserved as they fall towards each other (the motion
will be homothetic), so this must be a solution of the central limit
equation. For larger numbers of particles, there will be shell-like
structures in the solution. For even larger numbers, they will be
approximately spatially homogeneous \cite{BatGibSut03}.\newline

The mass $M$ of the system, given by (\ref{mass}), is not the same as the
constant  $\tilde{M}$, defined in (\ref{cosnsistency}), which
occurs crucially in the dynamical equations (\ref{ray}) and (\ref{Fried2})
below; the relation is to be investigated. Both are constant. If the masses $%
m_a$ are positive then in order to have solutions, $\tilde{M}$ must be
positive \cite{BatGibSut03}. This follows because on contracting (\ref{Newt3}%
) with $\mathbf{r}^a$,
\begin{equation}  \label{Newt3AAAA}
\tilde{M} = \sum_{b\neq a} m_b\frac{(\|\mathbf{r}_a\|^2-\mathbf{r}_b.\mathbf{%
r}^b)}{\|\mathbf{r}_a\|^2|\mathbf{r}_a-\mathbf{r}_b|^3}= \sum_{b\neq a} m_b%
\frac{(1- \cos^2\vartheta)}{|\mathbf{r}_a-\mathbf{r}_b|^3}
\end{equation}
where $\vartheta$ is the angle between $\mathbf{r}_a$ and $\mathbf{r}_b$.
Thus $\tilde{M}$ is uniquely determined by the mass distribution.\\

For approximately spherically symmetric distributions,
for which $m(r)$ is  the mass inside a radius $r$  we find,  by balancing forces
and using Newton's result that for a continuous spherical distribution of
mass, the gravitational field is given by the total mass inside a radius
$r$ divided by the radius squared,  we find
that
\begin{equation}
\tilde M = \frac{m(r)}{r^3} \,.
\end{equation}
It follows that for a spherical distribution, the density
of the central configuration  $\tilde \rho= \frac{3 \tilde M}{4 \pi}$
is homogeneous. The physical density $\rho$   is of course time dependent,
$\rho= \frac{3 \tilde M}{4 \pi S^3}$.\\

It would be possible to consider a spherically symmetric
density  distribution thought of as concentric shells
of density   $\rho(r)$ which would not be  homogeneous. That would  not satisfy
the condition for  a central configuration, but would
have a scale factor $S=S(t,r)$ which depends not only on time $t$
but radius $r$.

\subsubsection{A key identity}

\label{sec:idneitity} To obtain an integral relation following from the
central configuration  equation, multiply (\ref{Newt3}) by $G,$ take a dot product
with $\mathbf{r}_{a}$, and sum over $a$ 
to get
\begin{eqnarray}
G\tilde{M}\sum_{a}m_{a}\mathbf{r}_{a}.\mathbf{r}_{a}=\sum_{a}\sum_{b\neq a}
Gm_{a}m_{b}\frac{\mathbf{r}_{a}.(\mathbf{r}_{a}-\mathbf{r}_{b})}{|\mathbf{r%
}_{a}-\mathbf{r}_{b}|^{3}} = -\sum_{a}\sum_{b\neq
a}\mathbf{r}_{a}.\partial
_{\mathbf{r}_{a}}(\frac{Gm_{a}m_{b}}{|\mathbf{r}_{a}-\mathbf{r}_{b}|})
\,.\label{Newt31}
\end{eqnarray}%
Now Euler's theorem on homogeneous functions (in this case $\frac{1}{|%
\mathbf{r}_{a}-\mathbf{r}_{b}|}$ is of degree -1) shows%
\[
\mathbf{r}_{a}.\partial _{\mathbf{r}_{a}}(\frac{1}{|\mathbf{r}_{a}-\mathbf{r}%
_{b}|})=-\frac{1}{|\mathbf{r}_{a}-\mathbf{r}_{b}|}.
\]%
(cf. (\ref{euler1})), so the last term is
\begin{equation}  \label{Newt32}
-\sum_{a}\sum_{b\neq a}\mathbf{r}_{a}.\partial _{\mathbf{r}_{a}}(\frac{%
Gm_{a}m_{b}}{|\mathbf{r}_{a}-\mathbf{r}_{b}|}) = \sum_{a}\sum_{b\neq a}(%
\frac{Gm_{a}m_{b}}{|\mathbf{r}_{a}-\mathbf{r}_{b}|}) \,.
\end{equation}
Define the effective potential $\tilde{V}_{(-1)}$ and moment of
inertia $\tilde{I}_{0}$ by
\begin{eqnarray}
\tilde{V}_{(-1)}:= &&-\sum_{a}\sum_{b \neq a}\frac{Gm_{a}m_{b}}{|\mathbf{r}_{a}-\mathbf{r}%
_{b}|}=\mathbf{const}.  \label{V0} \\
\tilde{I}_{0}:=
&&\frac{1}{2}\sum_{a}m_{a}(\mathbf{r}_{a})^{2}=\mathbf{const}
\label{I0}
\end{eqnarray}%
(cf. (\ref{potdef}) and (\ref{cofm}); these are defined in terms of the comoving $\mathbf{r}_{a}$ rather than the
physical $\mathbf{x}_{a}$). Then on using (\ref{Newt32}), equation(\ref%
{Newt31}) becomes
\begin{equation}
2G\tilde{M}\,\tilde{I}_{0}=\ -\tilde{V}_{(-1)}  \label{virial}
\end{equation}%
showing that the central configuration moment of inertia $I_{0}$ and
effective potential energy $V_{(-1)}$ are equal up to a factor
$2G\tilde{M}$. This can be regarded as an analogue of the Virial
Theorem for static configurations. We will call it the \emph{central
configuration  constraint equation}, because it is a relation that is
required to be true if (\ref{Newt3}) is to hold for all $a$.\\

An alternative derivation of (\ref{virial})  is to substitute
the Slipher-Lema\^{i}tre-Hubble law (\ref{homothetic}) into
the Lagrange-Jacobi identity  (\ref{virialrelation}) and use
the Raychaudhuri and Friedmann equations (\ref{ray}), (\ref{Fried2}) below.

\subsubsection{The time evolution equations}

Secondly, from (\ref{defc}) and (\ref{cosnsistency}) we must have
\begin{equation}  \label{ray}
-\frac{G\tilde{M}}{S^2(t)} = \frac{d^2S(t)}{dt^2}
\end{equation}
which is the Raychaudhuri equation \cite{Ell71} for this case. One should
note that, for a given mass distribution, this equation is \emph{not}
invariant under rescaling $S(t) \rightarrow \hat{S}(t) = \alpha S(t)$; for
\begin{equation}  \label{ray33}
-\frac{G\tilde{M}}{\alpha^2 \hat{S}^2(t)} = \frac{d^2 (\alpha \hat{S}(t))}{%
dt^2}
\end{equation}
is the same as (\ref{ray}) only if the mass $\tilde{M}$ is rescaled also: $%
\tilde{M} \rightarrow \hat{\tilde{M}} = \alpha^3 \tilde{M}$. But for a given
mass distribution, $\tilde{M}$ is fixed by (\ref{Newt3AAAA}).\newline

Multiplying (\ref{ray}) by $(dS/dt)$, which must be non-zero for almost all $%
t$ because of (\ref{Newt1}), it can be integrated: 
\begin{equation}
\frac{d}{dt}\left( \frac{G\tilde{M}}{S(t)}\right) =-\frac{G\tilde{M}}{%
S^{2}(t)}\frac{dS(t)}{dt}=\frac{d^{2}S(t)}{dt^{2}}\frac{dS(t)}{dt}=\frac{1}{2%
}\frac{d}{dt}\left( \frac{dS(t)}{dt}\right) ^{2}  \label{integ1}
\end{equation}%
which gives the Friedmann equation 
\begin{equation}
\frac{G\tilde{M}}{S^{3}(t)}=\frac{1}{2}\left[ \frac{\dot{S}(t)}{S(t)}\right]
^{2}-\frac{E}{S^{2}(t)}  \label{Fried2}
\end{equation}%
where $E$ is a constant of integration. Thus we get the same result as both
the General Relativity and Newtonian cosmological continuum approximations
for the case of pressure free matter, but with no continuum model needed.

\subsubsection{Energy conservation}

How does this relate to the energy equation (\ref{energy})? They must both
represent the same process of energy conservation.\newline

Assuming the homothetic expansion hypothesis(\ref{homothetic}), the kinetic energy (\ref{energydef}) is
\[
T(\dot{\mathbf{x}_{c}})=\frac{1}{2}\sum_{a}m_{a}(\dot{\mathbf{x}}_{a})^{2}=%
\frac{1}{2}\dot{S}(t)^{2}\sum_{a}m_{a}(\mathbf{r}_{a})^{2}=\dot{S}%
^{2}(t)I_{0},
\]%
and the potential energy (\ref{potdef}) is
\begin{equation}
V(\mathbf{x}_{c})=-\sum_{a}\sum_{b \neq a}\frac{Gm_{a}m_{b}}{|\mathbf{x}_{a}-%
\mathbf{x}_{b}|}=-\frac{1}{S(t)}\sum_{a}\sum_{b \neq a}\frac{Gm_{a}m_{b}}{|\,.%
\mathbf{r}_{a}-\mathbf{r}_{b}|}=\frac{1}{S(t)}\tilde V_{(-1)}.
\end{equation}%
Thus the energy equation (\ref{energy}) is
\begin{equation}\label{energy111}
T+V=\dot{S}(t)^{2}\tilde{I}_{0}+\frac{1}{S(t)}\tilde{V}_{(-1)}=\mathcal{E}_{0}
\end{equation}%
which gives
\begin{equation}
\frac{1}{2}\frac{\dot{S}(t)^{2}}{S(t)^{2}}+\frac{1}{S^{3}(t)}\frac{\tilde{V}_{(-1)}}{%
2\tilde{I}_{0}}=\frac{\mathcal{E}_{0}}{2\tilde{I}_{0}}\frac{1}{S^{2}(t)}\,.
\end{equation}%
Comparing with (\ref{Fried2}), they agree if
\begin{equation}
E=\frac{\mathcal{E}_{0}}{2\tilde{I}_{0}},\,\,\,G\tilde{M}=-\frac{\tilde{V}_{(-1)}}{2\tilde{I}_{0}}
\end{equation}%
The former just relates the arbitrary constants $ E$ and
$\mathcal{E}_{0}$ and the latter is the central configuration constraint
equation (\ref{virial}). Thus equations (\ref{Fried2}) and (\ref{energy111}) are the same.

\subsubsection{The Virial Relation}
For a homothetic expansion (\ref{homothetic}), the moment of inertia (\ref{cofm}) becomes
\begin{equation}
I(t)=\frac{1}{2}\sum_{a}m_{a}x_{a}^{2}=S^{2}(t)\ \frac{1}{2}%
\sum_{a}m_{a}r_{a}^{2}=S^{2}(t)\tilde{I}_{0}  \label{mofIt}
\end{equation}
Taking a time derivative:
\begin{equation}\label{eq:dec}
\frac{dI(t)}{dt}=2\frac{dS(t)}{dt}S(t)\tilde{I}_{0}
\end{equation}
Take a second derivative:
\begin{equation}
\frac{d^{2}I(t)}{dt^{2}}=2\left(\frac{d^{2}S(t)}{dt^{2}}S(t)+
\dot{S}(t)^{2}\right)\tilde{I}_{0} \label{secondder}
\end{equation}%
Now (\ref{Fried2}) shows that
\begin{equation}\label{eq:decrr}
\dot{S}(t)^{2} = \frac{2G\tilde{M}}{S(t)}+ 2E
\end{equation}
Using this and (\ref{ray}), (\ref{secondder}) becomes
\begin{equation}
\frac{d^{2}I(t)}{dt^{2}}=2\left(-\frac{G\tilde{M}}{S^{2}(t)}S(t)+(\frac{2G\tilde{M}}{S(t)}+
2E)\right)\tilde{I}_{0}
=2(\frac{G\tilde{M}}{S(t)}+ 2E)\tilde{I}_{0}.
\end{equation}
This makes sense: as the system expands, the moment of inertia increases (cf. (\ref{eq:dec})) but at a decreasing rate (cf. (\ref{eq:decrr)}).
The virial relation (\ref{virialrelation}) becomes
\begin{equation}
V=2(\frac{G\tilde{M}}{S(t)}+ 2E)\tilde{I}_{0}-2T
\end{equation}
in contrast to the virial theorem (\ref{virialthm}). The condition $\langle \frac{d^{2}I(t)}{dt^{2}}\rangle = 0$ is not fulfilled.

\subsubsection{The main result}\label{sec:main_result}
These models can represent complex inhomogeneous matter distributions, but
not arbitrary ones. To summarise,

\begin{quote}
\textbf{Theorem: Discrete Newtonian Cosmology} \emph{The Newtonian
gravitational law of attraction (\ref{Newt}) for a finite set of gravitating
particles has an exact homothetic solution ((\ref{homothetic}), (\ref{vel})
hold) provided the time independent central configuration equation (\ref%
{Newt3}) is satisfied for $a = 1$ to $N$. The effect of gravitational
attraction is to lead to a homothetic change in size governed by the
Raychaudhuri equation (\ref{ray}), with first integral the Friedmann
equation (\ref{Fried2})}.
\end{quote}

These solutions are not spatially homogeneous (although they tend to spatial
homogeneity if the number of particles is large \cite{BatGibSut03}). Indeed
they break all the symmetries of the equations mentioned above. In
particular the origin of coordinates is a preferred point: it is the center
of mass (see (\ref{CofM}) below). Note that in the general relativity case, (%
\ref{ray}) and (\ref{Fried2}) imply each other because of the fluid energy
density conservation equation; essentially the same is true for Newtonian
fluid-based cosmology. Here the equivalence results in effect  from the mass
conservation equation (\ref{mass_cons}), which underlies the constancy of $%
\tilde{M}$.\newline

Roughly speaking, the central configuration equation (\ref{Newt3}) is the
condition that the matter distribution is homogeneous on a large scale,
allowing the quantity $M$ to be independent of spatial position. The effect
of gravitational attraction is to keep the spatial arrangement unchanged in
shape, but altering in size according to (\ref{ray}), (\ref{Fried2}); hence
in spatial terms, gravity leaves the configuration untouched apart from
homothetically altering distances.

\begin{quote}
\textbf{Corollary} \emph{There are no such FLRW-like solutions if the
central configuration equation (\ref{Newt3}) is not satisfied for all $a$ ($%
1\leq a\leq N$). The time development of data not satisfying these
conditions cannot be homothetic with a spatially homogeneous homothetic factor.%
}
\end{quote}
(cf. \cite{Arnold}: Proposition 2.5). If we relax the global homothetic assumption to a local self-similarity
condition:
\begin{equation}  \label{weaker}
\mathbf{x}_a = S(t,x)\mathbf{r}_a
\end{equation}
there will be many more solutions, as investigated by Saari \cite{Sar71}.


\subsection{Cosmological constant}\label{sec:cc}

\label{cc} The universe appears today
 to be dominated by a cosmological constant. Adding in a Newtonian cosmological constant to the force law, we get
\begin{equation}  \label{Newtlamb}
m_a \frac{d^2\mathbf{x}_a}{dt^2} = -\sum_{b\neq a} G m_a m_b\frac{(\mathbf{x}%
_a-\mathbf{x}_b)}{|\mathbf{x}_a-\mathbf{x}_b|^3} + \frac{\Lambda m_a \mathbf{%
x}_a}{3}
\end{equation}
As before, put in a homothetic factor and separate variables: using (\ref%
{homothetic}), 
(\ref{Newtlamb}) becomes
\begin{equation}  \label{Newtlamb1}
m_a \mathbf{r}_a\frac{d^2 S(t)}{dt^2} = -\sum_{b\neq a} G m_a m_b\frac{S(t)(%
\mathbf{r}_a-\mathbf{r}_b)}{S^3(t)|\mathbf{r}_a-\mathbf{r}_b|^3} + \frac{%
\Lambda S(t) m_a \mathbf{r}_a}{3}.
\end{equation}
The argument from (\ref{Newt1}) to (\ref{Newt3}) goes through as before.
This gives the result
\begin{equation}  \label{Newtlamb3}
m_a \mathbf{r}_a S^2(t) \frac{d^2 S(t)}{dt^2} = -G\tilde{M} m_a \mathbf{r}_a
+ \frac{\Lambda S^3(t) m_a \mathbf{r}_a}{3}
\end{equation}
with $\tilde{M}$ defined exactly as before by (\ref{Newt3}). This implies
the Raychaudhuri equation with cosmological constant:
\begin{equation}  \label{Newtlamb51}
\frac{1}{S(t)} \frac{d^2 S(t)}{dt^2} = -\frac{G\tilde{M}}{S^3(t)} + \frac{%
\Lambda}{3}
\end{equation}
where matter causes deceleration and $\Lambda$ an acceleration \cite{Ell71}.
To integrate when $dS/dt \neq 0$, multiply by $S(t)dS/dt$ to get
\begin{equation}  \label{integ11}
\frac{d^2S(t)}{dt^2}\frac{dS(t)}{dt} = -\frac{G\tilde{M}}{S^2(t)}\frac{dS(t)%
}{dt} + \frac{\Lambda}{3}S(t) \frac{dS(t)}{dt}
\end{equation}
which is
\begin{equation}  \label{integ12}
\frac{1}{2}\frac{d}{dt}\left(\frac{d S(t)}{dt}\right)^2 = \frac{d}{dt}\left(%
\frac{G\tilde{M}}{S(t)}\right) + \frac{d}{dt}\left(\frac{\Lambda S^2(t)}{6}%
\right).
\end{equation}
Integrating gives the Friedmann equation 
\begin{equation}  \label{Fried12}
\frac{1}{2}\left[\frac{\dot{S}(t)}{S(t)}\right]^2 = \frac{G\tilde{M}}{S^3(t)}
+ \frac{E}{S^2(t)} + \frac{\Lambda}{6}
\end{equation}
where $E$ is a constant of integration. The special case when $dS/dt =0$ is
dealt with below (Section \ref{sec:static}).

\begin{quote}
\textbf{Solutions with $\Lambda \neq 0$}: \emph{We can derive the standard
Raychaudhuri (\ref{Newtlamb51}) and Friedmann (\ref{Fried12}) equations for
time-dependent cosmology in exactly the same way for discrete Newtonian
cosmology with $\Lambda\neq 0$ as for the case with $\Lambda= 0$. The
central configuration equation (\ref{Newt3}) required for a homothetic
solution is unchanged (that equation does not gain a cosmological constant),
as is the definition of effective gravitational mass $\tilde{M}$. No fluid
approximation is used in deriving these results.}
\end{quote}

The possibility of such comoving homothetic configurations is not affected
 by  the cosmological constant,
which only affects the time evolution of the solution.
Other modifications of Newton's law of gravity,
such as those suggested by Neumann
\cite{Neumann} and by Seeliger \cite{Seeliger1,Seeliger2,Seeliger3,Seeliger4}
break  the scaling symmetry
of the inverse square law
and do not permit homothetic solutions (cf. \cite{Land1a}).\\

The introduction of the cosmological constant into (\ref{Newtlamb})
breaks the translation
symmetry of the original equations (\ref{Newt}) and one might
wonder about the fate of momentum conservation and the
 issue of the centre of mass. This is discussed in detail
in \cite{GibPat03} where it is explained how the Galilei invariance
of  (\ref{Newt}) is replaced by the Newton-Hooke invariance  of
(\ref{Newtlamb}).

\subsection{Specific Cosmological solutions}\label{sec:solutions}

Even though these discrete Newtonian solutions are spatially inhomogeneous,
their time dependence corresponds exactly to the pressure free-general
relativity models \cite{Ellvan98,Dod03,EllMaaMac12,PetUza13}.

\subsubsection{Static solutions}
\label{sec:static} In the case of static solutions ($S(t) = S_0$ = \emph{%
const}), (\ref{Fried12}) no longer follows from (\ref{Newtlamb51}), which is
the only gravitational equation to be satisfied apart from (\ref{Newt3}).
Solutions exist if and only if
\begin{equation}  \label{static1}
\frac{\Lambda}{3} = \frac{G\tilde{M}}{S_0^3}\,>0.
\end{equation}

\begin{quote}
\textbf{Static discrete mass solutions} \emph{exist for any central mass
configuration (\ref{Newt3}) provided $\Lambda > 0$. The only gravitational
equation to be satisfied in addition to (\ref{Newt3}) is (\ref{static1}),
with $\tilde{M} > 0$ defined by (\ref{Newt3AAAA}). Solutions clearly exist for
any values of $\tilde{M}$ and $\Lambda >0$ (one just has to solve (\ref{static1}) for $S_0$.)}
\end{quote}

These are discrete Newtonian analogues of the Einstein static solution; just
as in the general relativity case, they will be unstable \cite{Ell71}. There
may be no General Relativity analogues of these static discrete mass
solutions \cite{UzaEllLar11}.

\subsubsection{Expanding solutions}

The dynamic models with $\Lambda > 0$ can be good descriptions of the real
universe after the universe is matter dominated, and specifically since the
time of decoupling of matter and radiation \cite{Ellvan98,Dod03,EllMaaMac12,PetUza13}. \newline

These solutions depend in the standard way on $\tilde{M}$, $E$, and $\Lambda$%
, allowing monotonic solutions only if $E \geq 0$, and a much wider set of
solutions otherwise \cite{EllMaaMac12,PetUza13}. Assuming $\Lambda \geq 0$,
bounces can occur if and only if $E < 0$; otherwise the universe had a
singular start where $S(t) \rightarrow 0$. The universe will expand forever
unless $E<0$, when it may recollapse. Exact parametric solutions can be
obtained when $\Lambda=0$ \cite{Ellvan98}. The simplest solution is the
Einstein-de Sitter solution, arising when $\Lambda = 0$, $E = 0$, leading to
\begin{equation}  \label{Eds}
S(t) = S_0 t^{2/3}\,,\,\,S_0 = \left(\frac{9}{2}G\tilde{M}\right)^{1/3}\,,
\end{equation}
which gives a solution for any $\tilde{M}$ (unlike the general relativity
case, we do not have the freedom to rescale $S$). This solution corresponds
to the quantity $\mathbf{X}_a = t^{-2/3} \mathbf{x}_a$ being constant (see (%
\ref{invcon})). \newline

Asymptotic solutions for large $t$ (in the cases that expand forever) fall
into three cases: 1. $\Lambda>0$, 2. $\Lambda=0$, $E > 0$, 3. $\Lambda=0$, $%
E = 0$. In case 1., the asymptotic 
solution is the de Sitter solution
\begin{equation}  \label{Fried12345}
S(t) = \alpha \exp Ht,\, H = \left(\frac{\Lambda}{3}\right)^{1/2}\,,
\end{equation}
where $\alpha$ is arbitrary (the solution is scale free). In case 2., the
asymptotic 
solution is the Milne solution
\begin{equation}  \label{Fried99}
S(t) = H t,\, H^2 = 2 E,
\end{equation}
which is again independent of $\tilde{M}$. In case 3., the asymptotic
solution is the same as the exact solution (\ref{Eds}). The last two
solutions are consistent with Saari's analysis of asymptotic forms \cite%
{Sar71} (which did not consider the case $\Lambda > 0$).\newline

These models do not represent well the dynamics of the universe at
early times when radiation dominates and general relativity effects
therefore have to be taken into account, both because the active
gravitational mass then is $(\rho +3p/c^2)$ rather than $\rho$, and
because the energy conservation equation has a source term $(\rho +
p/c^2)$ rather than $\rho$ \cite{Ell71}. However structure formation
takes place after decoupling; and so these equations may be good at those
times. In order to investigate this we need to look at the perturbed
equations (these are to be the topic of a subsequent paper).

\subsection{Issues}
\label{sec:issues} For the FLRW type situation defined by (\ref{homothetic}),
we have an intriguing fine tuning problem:

\begin{quote}
\textbf{Fine Tuning}: \emph{To good approximation, we currently see a FLRW
type homothetic expansion. But in order to get such a flow, the initial
positions of the particles must be constrained to satisfy (\ref{Newt3}).
What kind of explanation can one give for such a fine tuning of the initial
data in Newtonian cosmology}?
\end{quote}

Presumably the answer is to be sought via an initial relativistic state that
results at late times in such a Newtonian configuration. Perhaps also it can
be justified by a minimum energy principle favouring this distribution, but
it is not clear how this might work:
\begin{itemize}
\item Every central configuration is an extremum,

\item but not every extremum is a minimum (or maximum).
\end{itemize}
(see \cite{BatGibSut03}, section 2(b)).


\section{The central configuration equation}
\label{sec:cce}
As shown above, the central configuration equation (\ref{Newt3}):
\begin{equation}  \label{Newt3A}
\tilde{M} m_a \mathbf{r}_a = \sum_{b\neq a} m_a m_b\frac{(\mathbf{r}_a-%
\mathbf{r}_b)}{|\mathbf{r}_a-\mathbf{r}_b|^3}
\end{equation}
is the initial value equation for discrete Newtonian cosmology; once it has
been satisfied at an initial time, it will be satisfied for all times (that
is the essence of (\ref{cosnsistency})). It plays a key role in celestial
mechanics \cite{Sar71,Mar76,Sar80} and its solutions have been studied
in depth in \cite{BatGibSut03}, but deserve much more study.\newline

Considered as a 3-dimensional gravitational problem, it is as if
there were a force proportional to distance between the particles,
as well as the inverse square law of attraction; that is, it is as
if there were a cosmological constant in this 3-dimensional of
gravitational context. But it is not the same as a cosmological
constant $\Lambda$ (see previous section), which has no effect on
the spatial gravitational attraction equation (\ref{Newt3A}); rather
$\Lambda$ changes the time evolution of the system, see
(\ref{Fried12}). One may think of a central configuration as an
equilibrium between the gravitational attraction and an entirely
fictitious or auxiliary cosmological repulsion which arises \' a la
D'Alembert's principle from the inertial forces due to the
accelerations of the particles. Thus this is an effective
3-dimensional force arising for the 4-dimensional  spacetime context (see
Section \ref{effective} for further discussion).

\subsection{Some Properties of Central Configurations}

This section follows section 2 of \cite{BatGibSut03}.

\subsubsection{Centre of mass}

The centre of mass $\mathbf{r}_{CM}$ is given by
\begin{equation}  \label{CofM}
M \mathbf{r}_{CM} = \sum_a m_a \mathbf{r}_a = \sum_a \sum_{b\neq a} \frac{%
m_a m_b}{\tilde{M}}\frac{(\mathbf{r}_a-\mathbf{r}_b)}{|\mathbf{r}_a-\mathbf{r%
}_b|^3} = 0
\end{equation}
because the sum is symmetric but the summand antisymmetric. Thus the
centre of mass of the system lies at the origin, which is a
preferred location for these inhomogeneous distributions.
In this model therefore, Neumann's body alpha (some sort of fixed body defining inertial motion),
which played an important role in the pre-relativity debate
about absolute versus relative motion in Newtonian mechanics \cite{Neumann,Giorgi,Whittaker},
may be identified with the origin.\\

Taking the
time derivative
\begin{equation}  \label{CofM1}
\mathbf{P} := \sum_a m_a \dot{\mathbf{x}}_a = \dot{S}(t) \sum_a m_a
\mathbf{r}_a = 0
\end{equation}
so the conserved momentum is zero. Similarly for angular momentum about the
centre of mass:
\begin{equation}  \label{AM1}
\mathbf{L} = \sum_a m_a (\mathbf{x}_a \times \,\dot{\mathbf{x}}_a) =
S(t) \dot{S}(t) \sum_a m_a (\mathbf{r}_a \times\,
\dot{\mathbf{r}}_a) = 0.
\end{equation}

Solutions with vanishing total  momentum and total angular momentum
are sometimes referred to as  ``relational''.
For an illuminating  discussion of the relation of this to various formulations
of Mach's Principle the reader is referred to
\cite{Zanstra,Ding2,Ding3,Bondi2,BarPfi95}.

\subsubsection{Effective Forces}
\label{effective} 
One can represent the nature of the central configuration in terms of effective forces and potentials (effective because they refer to the comoving distances
$\mathbf{r}_a$ rather than the actual distances $\mathbf{x}_a$ that occur in the underlying force equation (\ref{grav}) and its resulting potentials
(\ref{potdef1}).\\

Starting with (\ref{mass}), 
add and subtract the same term to get
\begin{equation}  \label{mass2}
m_a \mathbf{r}_a = \frac{1}{M}m_a \mathbf{r}_a \sum_b m_b = \frac{1}{M}
\sum_b m_bm_ a (\mathbf{r}_a - \mathbf{r}_b) + \frac{m_a}{M} \mathbf{r}_a
\sum_b m_b \mathbf{r}_b.
\end{equation}
Using (\ref{CofM})
\begin{equation}  \label{mass22}
m_a \mathbf{r}_a = \frac{1}{M} \sum_{b\neq a} m_bm_ a (\mathbf{r}_a
- \mathbf{r}_b).
\end{equation}
Substituting this into (\ref{Newt3A}) and multiplying by $G$, we get
\begin{equation}  \label{Newt3B}
\sum_{b\neq a} G m_ bm_a (\mathbf{r}_a-\mathbf{r}_b) \left(\frac{G\tilde{M}}{%
M} - \frac{G}{|\mathbf{r}_a-\mathbf{r}_b|^3}\right) = 0
\,.\end{equation}
Defining $r_{ab} :=|\mathbf{r}_a -\mathbf{r}_b|$ and the effective
inter-particle force
\begin{equation}  \label{Forceinter}
\mathbf{\tilde{F}}_{ab}:= m_ bm_a (\mathbf{r}_a-\mathbf{r}_b) \left(\frac{G%
\tilde{M}}{M} - \frac{G}{r_{ab}^3}\right)
\end{equation}
we find that (\ref{Newt3B}) is just
\begin{equation}  \label{Force1}
\sum_{a\neq b} \mathbf{\tilde{F}}_{ab} = 0.
\end{equation}
which is a form that is invariant under translation of the points: the
centre of mass does not matter.\newline

We can rewrite (\ref{Forceinter}) as
\begin{equation}  \label{Force2}
\mathbf{\tilde{F}}_{ab} = \mathbf{F}_{ab}^{(TD)} + \mathbf{\tilde{F}}%
_{ab}^{(1)}
\end{equation}
where
\begin{equation}  \label{fgrav}
\mathbf{\tilde{F}}_{ab}^{(1)} := - G m_a m_b\frac{(\mathbf{r}_a-\mathbf{r}_b)}{%
|\mathbf{r}_a-\mathbf{r}_b|^3}
\end{equation}
is the reduced inter-particle gravitational force, which relates to the
proper distances $\mathbf{x}_\alpha$ rather than the comoving distances $%
\mathbf{r}_\alpha$ (cf.(\ref{grav})), and
\begin{equation}  \label{Force3}
\mathbf{F}_{ab}^{(TD)} := G\left(\frac{\tilde{M}}{M}\right) m_ bm_a (\mathbf{%
r}_a-\mathbf{r}_b)
\end{equation}
is the \emph{top-down (contextual) effective force} exerted on the spatial
distribution because of the context of the conformal expansion. It is an
effective repulsive force that is obviously not the same as the direct
gravitational force between the particles, since  that  given by $\mathbf{F}%
_{ab}^{(grav)}$. It is not due to a cosmological constant (cf. Section \ref%
{cc}); it is an extra term that arises solely due to the configuration of
particles, rather than the micro forces between them. This is in line with
many other examples of such contextual effects in physics \cite{Ell12}.%
\newline

It follows that the effective inter-particle force vanishes when
\begin{equation}  \label{Force5}
\mathbf{\tilde{F}}_{ab}= 0 \Leftrightarrow |\mathbf{r}_{ab}| = R_c := \left(%
\frac{G M}{\tilde{M}}\right)^{1/3}
\end{equation}
giving a preferred scale for these solutions. This is discussed further in
\cite{BatGibSut03}.

\subsection{Potential Functions}
\subsubsection{Potentials for particles}
\label{sec:potential} Write the central configuration equation
(\ref{Newt3A}) as
\begin{equation}  \label{forceplus}
\mathbf{\tilde{F}}_a := \mathbf{\tilde{F}}_{a}^{(1)} + \mathbf{\tilde{F}}_{a}^{(2)}
= 0\,
\end{equation}
where $\mathbf{\tilde{F}}_a^{(1)}$ is given by
\begin{equation}  \label{fgrav11}
\mathbf{\tilde{F}}_a^{(1)}=  \sum_{b\neq a}\mathbf{\tilde{F}}_{ab}^{(1)}
= - \sum_{b\neq a} G m_a m_b\frac{(\mathbf{r}_a-\mathbf{r}_b)}{%
|\mathbf{r}_a-\mathbf{r}_b|^3}
\end{equation}
on using (\ref{fgrav}); and $\mathbf{\tilde{F}}_a^{(2)}$ is defined
by
\begin{eqnarray}
\mathbf{\tilde{F}}_a^{(2)}: = G \tilde{M} m_a \mathbf{r}_a.
\end{eqnarray}
Note that these are defined in terms of the comoving coordinates $\mathbf{r}%
_a$ rather than the Newtonian coordinates $\mathbf{x}_a$ Define the
associated energies as
\begin{equation}  \label{potentialy}
\tilde{V}_a := \tilde{V}_{(-1)a} + \tilde{V}_{(2)a}
\end{equation}
where the effective gravitational potential energy is 
\begin{equation}  \label{potential11y}
\tilde{V}_{(-1)a} := - \sum_{b \neq a} \frac{G m_a
m_b}{|\mathbf{r}_{ab}|}
\end{equation}
which is homogeneous of degree $-1$, and the effective repulsion potential
energy is
\begin{equation}  \label{potential2}
\tilde{V}_{(2)a} := -\frac{1}{2} G \tilde{M}  m_a \textbf{r}_a.\textbf{r}_a
\end{equation}
which is homogeneous of degree $k=2$. These are also defined in terms of the
comoving coordinates $\mathbf{r}_a$.\\

From these definitions,
as in the case of (\ref{pot1}),
\begin{equation}\label{popot}
\mathbf{\tilde{F}}_{a}^{(1)} = -\frac{\partial \bar{V}_{(-1)a}}{\partial \mathbf{r}_{a}},\,\,
\mathbf{\tilde{F}}_{a}^{(2)} = - \frac{\partial  \bar{V}_{(2)a}}{\partial \mathbf{r}_{a}}
\end{equation}
Solutions of the central configuration
equation are critical points of $\tilde{V}_a$:
\begin{equation}  \label{critical}
\mathbf{\tilde{F}}_a = 0 \Leftrightarrow \frac{\partial \tilde{V}_a}{\partial
\mathbf{r}_{a}} = \frac{\partial }{\partial \mathbf{r}_{a}}\tilde{V}_{(-1)a} + \frac{%
\partial }{\partial \mathbf{r}_{a}}\tilde{V}_{(2)a}  = 0.
\end{equation}

\subsection{A Variational Principle for Central Configurations}
Define the
associated total energies as
\begin{equation}  \label{potentialz}
\tilde{V} := \tilde{V}_{(-1)} + \tilde{V}_{(2)}
\end{equation}
where the effective total gravitational potential  energy is 
\begin{equation}  \label{potential11y11}
\tilde{V}_{(-1)} := \sum_a \tilde{V}_{(-1)a} = - \sum_a \sum_{b \neq
a} \frac{G m_a m_b}{|\mathbf{r}_{ab}|}
\end{equation}
which is homogeneous of degree $-1$, and the effective total repulsion potential
energy is
\begin{equation}  \label{potential2zz}
\tilde{V}_{(2)} := \sum_a \tilde{V}_{(2)a} = -\frac{1}{2} \sum_a G \tilde{M}  m_a \textbf{r}_a.\textbf{r}_a
\end{equation}
which is homogeneous of degree $k=2$. These are also defined in terms of the
comoving coordinates $\mathbf{r}_a$. By definition, they are all constant.\\

Now (\ref{potential2zz}) is just $\tilde{V}_{(2)} := - G \tilde{M} \tilde{I}_0$ and (\ref{virial}%
) is $2G\tilde{M}\,\tilde{I}_{0}=\ -\tilde{V}_{(-1)} $, so together
they give the virial-type relation
\begin{equation}  \label{potential3}
\tilde{V}_{(-1)} = 2 \tilde{V}_{(2)}.
\end{equation}
for the effective energies (cf. equation (\ref{virialthm})).
This implies that one can express the total effective energy in terms of either partial term:
\begin{equation}  \label{potential5}
\tilde{V} := \frac{3}{2}\tilde{V}_{(-1)} = 3 \tilde{V}_{(2)}
\,.\end{equation}
However as noted above, the virial theorem does not hold for the space-time system.

Critical points of  $\tilde{V}_a$ are clearly critical points of
$\tilde{V} = \sum_a \tilde{V}_a$. Since $\tilde V$ becomes
infinitely large and negative as two or more points approach one another
or as one or more recede to infinity,  it is easy to see that
there must be at least one global maximum and no global minimum.
In addition one suspects there are many saddle points.
Thus we have

\begin{quote}
\textbf{Theorem:} \emph{Critical points of the
function $\tilde V $ are in one-one correspondence with central configurations.
There is at least one  global maximum  and no global minimum.
Every critical point, and hence every central configuration,
satisfies $2 \tilde V_{2} = \tilde V_{(-1)} $.}
\end{quote}

The importance of this result is that it allows one
to search efficiently for   central configurations of very
many particles by numerically maximising $\tilde V$ \cite{BatGibSut03}.\\

\emph{ It is important to realise  that whether or not the
critical point is a maximum or otherwise of $\tilde V$
 is unrelated to whether
of not the associated time dependent solution of (\ref{Newt}) is
dynamically stable.}


\subsection{A scale-free variational principle for Central Configurations}

We  showed in the previous subsection  that every  central configuration is
a critical point of the function $ \tilde V= \tilde V_{(-1)} + \tilde V_{(2)}$
which is defined  on the $3N$-dimensional configuration space $Q_N$ of
$N$ distinct points  in three-dimensional  Euclidian space.
Moreover  every critical point satisfies

\begin{equation} \tilde V_{(-1)} =2 \tilde V_{(2)} .
\label{VR} \end{equation}
An equivalent formulation, suggested to us by the work of Julian Barbour (private communication),
is obtained by considering the scale-invariant function
\begin{equation}
C_s= -(- \tilde V_{(2)} ) ^{1/2} \tilde V_{(-1)} = - I^{1/2} V
\end{equation}
which is defined on the quotient $Q_N/{\Bbb R}_+$   of $Q_N$ by homotheties.
Because of the scaling properties of these quantities, replacing
$\textbf{x}_a$ by $\textbf{r}_a = \textbf{x}_a/S(t)$ leaves $C_s$ unchanged
\begin{equation}  \label{cs1}
C_s =  - \tilde{I}_{0}^{1/2}\, \tilde{V}_{(-1) \,.}
\end{equation}

If we differentiate $C_s$ with respect to $\mathbf{r}_a$ we obtain
\begin{equation}
-\frac{1}{2} (- \tilde V _{(2)} ) ^{-{1/2} }\tilde  V_{(-1)}
 \frac{\partial  \tilde V _{(2)} }{\partial \mathbf{r}_a} +
 (- \tilde V _{(2)} ) ^{{1/2} }  \frac{\partial  \tilde V _{(-1)} }
{\partial \mathbf{r}_a} =0 \,. \label{scale}
\end{equation}
Taking the dot product with $\mathbf{r}_a $, summing over $a$
and using Euler's theorem gives (\ref{VR})
and substituting back into (\ref{scale})  yields
\begin{equation}
  \frac{\partial  \tilde V _{(2)} }{\partial \mathbf{r}_a}
+\frac{\partial  \tilde V _{(-1)} }
{\partial \mathbf{r}_a} =  \frac{\partial  \tilde V }{\partial \mathbf{r}_a} =0\,. \label{critical2}
\end{equation}
Thus every critical point of $C_s$ is a critical point of $\tilde V$
and moreover satisfies (\ref{VR}).
Conversely (\ref{critical2})  and (\ref{VR} )  are easily seen to imply
(\ref{scale}). Thus we have the following

\begin{quote}
\textbf{Theorem:}: \emph{Critical points of the scale-invariant
function $C_s$ are in one-one correspondence with central configurations.}
\end{quote}

We repeat the warning that \emph{it is important to realise  that whether or not the
critical point is a maximum or otherwise of $C_s$ is unrelated to whether
of not the associated time dependent solution of (\ref{Newt}) is
dynamically stable.}



\section{Elaborations}\label{sec:further}

This paper has set out how exact Newtonian solutions exist for homothetically moving configurations of gravitating point particles. One can consider changes to the problem if there are the following generalisations:
\begin{itemize}
\item An environment of objects that are not affected by the system,
exerting an external gravitational field,
\item Subgroups of particles that are distinguished by being held together
by elastic forces,
\item If the  size of particles is too large for them to be considered as points, so
tidal forces matter,
\item Inhomogeneous conformal solutions,
\item Solutions with rotation as well as expansion.
\end{itemize}
While these are all of interest, the key further development is to consider the perturbed version of these equations, and how they relate to structure formation after decoupling. That will be the topic of a further paper.\\

We thank Roy Maartens, John Barrow, and especially Jeremy Butterfield for useful comments.


\end{document}